# Energy Efficiency in Hybrid Beamforming Large-scale mmWave Multiuser MIMO with Spatial Modulation

Merve Yüzgeçcioğlu
Communications Laboratory
TU Dresden
D-01062 Dresden, Germany
merve.yuzgeccioglu@tu-dresden.de

Alessio Zappone
Large Networks and Systems Group
L2S - CentraleSupelec, CNRS, Univ. Paris Sud,
Gif-Sur-Yvette, France
alessio.zappone@l2s.centralesupelec.fr

Eduard Jorswieck
Communications Laboratory
TU Dresden
D-01062 Dresden, Germany
eduard.jorswieck@tu-dresden.de

*Abstract*—The problem of radio resource allocation for global energy efficiency (GEE) maximization in mmWaves large-scale multiple-input multiple-output (MIMO) systems using hybrid-beamforming with spatial modulation is addressed. The theoretical properties of the optimization problem at hand are analyzed and two provably convergent optimization algorithms with affordable complexity are proposed. The former achieves the global optimum, while the latter trades off optimality with a lower computational complexity. Nevertheless, numerical results show that both algorithms attain global optimality in practical scenarios.

## I. Introduction

The number of connected devices is increasing drastically [1] and requires a 1000x data rate increase for future networks. However, this goal can not be achieved by simply scaling up the transmit power due to cost and energy consumption reasons. Therefore, bit-per-Joule energy efficiency has emerged as an enabling performance metric for future networks [2] and many research efforts have been conducted to develop radio resource allocation algorithms for energy efficiency maximization in various instances of cellular networks [3]–[6], and references therein. These works establish that considerable energy savings can be achieved by a suitable power allocation policy.

In order to meet the demand of 5G and beyond 5G networks, a very promising technology is the use of large-scale multiple-input multiple-output (MIMO) [7]–[9] systems in the mmWaves range [10], [11]. However, in large-scale mmWave MIMO, it is not possible to equip each antenna with a dedicated RF chain due to cost and energy consumption. Therefore, hybrid analog and digital beamformers must be employed [12]–[14]. In [15], a near-optimal hybrid beamforming structure is proposed in a multi-user scenario in order to mitigate inter-user-interference. Another approach is introduced in [16], where a fully-connected structure with minimum number of RF chains is developed. The singular vectors of the channel matrix are exploited in [17] to generate hybrid beamformer with lower complexity.

Another recent trend in order to reduce energy consumption, is the use of spatial modulation (SM) [18], [19], in systems with hybrid beamforming. From an energy-efficient perspective, SM is attractive because it enables to transmit additional information bits besides conventionally-modulated symbols, without requiring extra power [18], [20], [21]. SM in multi-user MIMO networks is addressed in [22] and a precoding method is proposed in order to cancel inter-user-interference. Combination of SM and beamforming is investigated in [23], where analog beamforming is employed to develop a generalized modulation scheme for single-user systems impaired by Rician fading. A closer look into the transmitter design of spatial modulators using analog beamforming is given in [24]. Furthermore, SM with hybrid beamforming has been proposed for a mmWave railway communication system in [25].

Nevertheless, all above works focus on optimizing the data rate of SM systems, whereas the problem of radio resource allocation for bit-per-Joule energy efficiency maximization has never been studied in large-scale MIMO systems employing SM and hybrid beamforming. In this context, an important point to stress is that, unlike conventional systems in which energy-efficient radio resource allocation basically amounts to tuning the transmit power levels, in large-scale MIMO systems the number of active RF chains is also an extremely important metric to optimize, because the system hardware power consumption is proportional to the antenna number [26]. A recent contribution in this sense is [27], where sequential optimization is used to determine the antenna number that maximizes the energy efficiency of a single wireless backhaul link. In [14], different hybrid receiver architectures are compared with respect to their energy efficiency and channel estimation performance. However, no radio resource allocation algorithm is derived.

Motivated by this background, this work proposes a novel optimization framework for bit-per-Joule energy efficiency maximization in the downlink channel of a mmWave large-scale MIMO system with SM and hybrid beamforming. In this scenario, we show that the number of active RF chains is related to the number of active users and the resource allocation problem is formulated as the maximization of the system energy efficiency with respect to the transmit power and number of served users, subject to maximum power constraints and minimum and maximum number of active users. First, the optimal number of users for any fixed power level, and the optimal power level for any fixed number of users are derived in closed-form. Next, two optimization algorithms are proposed to jointly allocate the users' number and the transmit power. Both algorithms are provably convergent with affordable complexity. The former is globally optimal, while the latter enjoys an even lower complexity, but is not

The work of Merve Yüzgeçcioğlu has received partly funding from the European Union's Horizon 2020 research and innovation programme under the Marie Sklodowzka-Curie grant agreement No 641985. The Work of A. Zappone has received funding from the H2020-MSCA-IF-2016 BESMART, under grant Grant 749336.

provably optimal. Numerical algorithms show that in practical scenarios both algorithms achieve global optimality.

## II. SYSTEM MODEL

The considered downlink system is depicted in Fig. 1. The base station (BS) is equipped with $N_A$ antenna arrays (AA), each having $N_T$ transmit antennas without any common antenna between different AAs.[1] Moreover, $N_{RF}$ RF chains are deployed at the BS to serve $K$ users. Each user has $N_R$ receive antennas and each receive antenna is connected to an RF chain. Additionally, the transmitter employs SM. That means that the incoming bits intended for a specific user are divided into two parts: the former is modulated by choosing one of the AAs, thus allowing the transmission of $\log_2 N_A$ bits per user. The latter are modulated by a conventional $M$-ary modulation scheme, thereby enabling the transmission of $\log_2 M$ bits. The total number of bits transmitted by the BS at each channel use is $K(\log_2 N_A + \log_2 M)$. According to the incoming bits for each bit stream one of the AA indices is chosen and the $M$-ary modulated data is transmitted by using the selected AA. Since every AA represents an SM symbol, it is possible that multiple users are served simultaneously by one AA. The steering of the beams towards the intended users is performed by the analog beamformers.

The number of data streams the BS has to process is the number of users $K$. Since this represents a lower bound on the number of RF chains $N_{RF}$, it is convenient to set $N_{RF} = K$. With the described notation, the received signal of the $i$-th user is written as

$$\mathbf{r}_i = \sqrt{\rho} \sum_{j=1}^{K} \mathbf{H}_{a_j,i} \mathbf{f}_{a_j,j} \mathbf{p}_{a_j,j} \mathbf{s} + \mathbf{n}_i, \quad (1)$$

Herein, $\mathbf{H}_{a_j,i} \in \mathbb{C}^{N_R \times N_T}$ is the $L$-path channel between the $a_j$-th AA and $i$-th user where $a_j$ is the selected AA index to transmit data to $j$-th user. The channel follows a geometry-based model shown below [28]

$$\mathbf{H}_{a_j,i} = \sqrt{\frac{N_T N_R}{L}} \sum_{l=1}^{L} \alpha_{a_j,i_l} \mathbf{a}_{R_{a_j,i}}(\theta_l) \mathbf{a}_{T_{a_j,i}}^H(\phi_l), \quad (2)$$

Here, $\mathbf{a}_{R_{a_j,i}}(\theta_l) = \sqrt{\frac{1}{N_R}}[1, e^{j\pi \sin \theta_l}, \ldots, e^{j\pi(N_R-1)\sin\theta_l}]^T$ and $\mathbf{a}_{T_{a_j,i}}(\phi_l) = \sqrt{\frac{1}{N_T}}[1, e^{j\pi \sin \phi_l}, \ldots, e^{j\pi(N_T-1)\sin\phi_l}]^T$ are the receive and transmit AA responses of the $l$-th path where $l = 1, \ldots, L$ with uniform linear array (ULA) structure. $\phi_l$ and $\theta_l$ are the angle of departure (AoD) and the angle of arrival (AoA) of the path and drawn from the uniform distribution $\mathcal{U}(0, 2\pi]$. Finally, $\alpha_{a_j,i_l}, \mathcal{CN}(0,1)$, represents the channel gain and the path loss of the path. The analog beamformer and digital precoder vectors of user $j$ are $\mathbf{f}_{a_j,j} \in \mathbb{C}^{N_T \times 1}$ and $\mathbf{p}_{a_j,j} \in \mathbb{C}^{1 \times K}$, which are considered fixed and chosen as described in [29]. The vector $\mathbf{s} \in \mathbb{C}^{K \times 1}$ contains the symbols of all the users, and finally $\mathbf{n}_i \in \mathbb{C}^{N_R \times 1}$ is the noise vector, modeled as a complex Gaussian vector with zero-mean and covariance matrix $\sigma^2 \mathbf{I}_{N_R}$. After receive combining, the received data vector is expressed as

$$y_i = \mathbf{w}_{a_i,i}^H \sqrt{\rho} \sum_{j=1}^{K} \mathbf{H}_{a_j,i} \mathbf{f}_{a_j,j} \mathbf{p}_{a_j,j} \mathbf{s} + \mathbf{w}_{a_i,i}^H \mathbf{n}_i, \quad (3)$$

[1] All results to follow could be easily extended to the case of different antenna number per AA.

where the receive combiner $\mathbf{w}_{a_i,i}$ is chosen according to [29], and $\rho$ is the total power transmitted by the BS. Assuming uniform power allocation, the transmit power allocated to each user is $\rho_i = \rho/K$, and since the interference from other users is eliminated by employing transmit precoding, (3) simplifies to

$$y_i = \sqrt{\rho_i} \mathbf{w}_{a_i,i}^H \mathbf{H}_{a_i,i} \mathbf{f}_{a_i,i} s_i + \mathbf{w}_{a_i,i}^H \mathbf{n}_i. \quad (4)$$

The mutual information between transmitted and received symbols is $I(y_i; x_i) = h(y_i) - h(z_i)$ with $x_i$ the symbol that contains the AA index and $M$-ary modulated symbol, $z_i \sim \mathcal{CN}(0, \sigma^2)$ the noise term after receive combining, while $y_i$ follows a Gaussian mixture distribution

$$f_{Y_i}(y_i) = \frac{1}{MN_A} \sum_{m=1}^{M} \sum_{j=1}^{N_A} \frac{1}{\pi} \exp\left\{-\frac{|y_i - \sqrt{\rho_i} \mathbf{w}_{j,i}^H \mathbf{H}_{j,i} \mathbf{f}_{j,i} s_m|^2}{\sigma^2}\right\}, \quad (5)$$

wherein $\mathbf{H}_{j,i}$ is the channel matrix of the $j,i$ antenna array-user pair, $s_m$ denotes the $m$-th symbol from the $M$-ary constellation diagram.

The fact that (5) follows a Gaussian mixture distribution complicates the derivation of an achievable rate expression, and indeed an exact closed-form rate expression is currently not available in the literature. Moreover, the analysis is complicated by the fact that discrete symbols from an $M$-ary constellation are considered here. In order to proceed further, we derive a tractable upper-bound of user $i$'s achievable rate in the next proposition.

**Proposition 1.** *For all $i$, the achievable rate $R_i$ of user $i$ assuming discrete-symbol input, is upper-bounded as*

$$R_i \leq \log_2 \pi e \left[ \sigma^2 + \frac{\rho}{KMN_A} \sum_{m=1}^{M} \sum_{j=1}^{N_A} |\mathbf{w}_{j,i}^H \mathbf{H}_{j,i} \mathbf{f}_{j,i} s_m|^2 \right.$$
$$- \frac{\rho}{KM^2N_A^2} \left(\sum_{m=1}^{M} \sum_{j=1}^{N_A} \Re\{\mathbf{w}_{j,i}^H \mathbf{H}_{j,i} \mathbf{f}_{j,i} s_m\}\right)^2$$
$$\left. - \frac{\rho}{KM^2N_A^2} \left(\sum_{m=1}^{M} \sum_{j=1}^{N_A} \Im\{\mathbf{w}_{j,i}^H \mathbf{H}_{j,i} \mathbf{f}_{j,i} s_m\}\right)^2 \right]$$
$$- \log_2 \pi e \sigma^2 = \bar{R}_i \geq 0. \quad (6)$$

*Proof:* The proof is omitted due to space constraints. It can be found in [30]. ∎

Elaborating, the right-hand-side of (6) can be expressed as

$$\bar{R}_i = \log_2\left(1 + \frac{\rho}{K} a_i\right), \quad (7)$$

with

$$a_i = \sum_{m=1}^{M} \sum_{j=1}^{N_A} \frac{|b_{j,m,i}|^2}{MN_A}$$
$$- \left(\sum_{m=1}^{M} \sum_{j=1}^{N_A} \frac{\Re\{b_{j,m,i}\}}{MN_A}\right)^2 - \left(\sum_{m=1}^{M} \sum_{j=1}^{N_A} \frac{\Im\{b_{j,m,i}\}}{MN_A}\right)^2, \quad (8)$$

and $b_{j,m,i} = \frac{\mathbf{w}_{j,i}^H \mathbf{H}_{j,i} \mathbf{f}_{j,i} s_m}{\sigma}$.

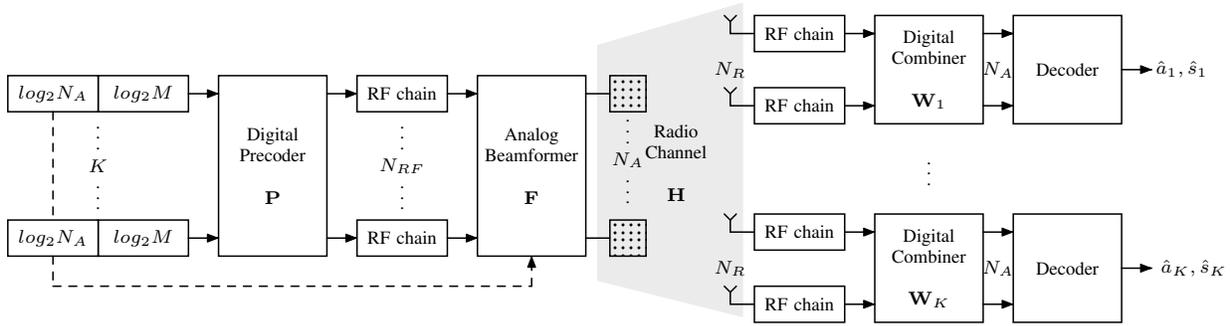

Fig. 1: Block diagram of hybrid beamforming with spatial modulation in multi-user downlink transmission

### III. Energy Efficiency Optimization

This section formulates the energy efficiency maximization problem and develops two solution algorithms.

#### A. Problem formulation

The ergodic sum-rate of the considered network is expressed as

$$R_s(K,\rho) = \sum_{i=1}^{K} \mathbb{E}\left[R_i(K,\rho)\right] = K\mathbb{E}\left[R_1(K,\rho)\right], \quad (9)$$

wherein it has been exploited the fact that the random variables $R_i$, $i = 1,\ldots,K$, are identically distributed. On the other hand, the total network power consumption is

$$P_T(K,\rho) = \rho + KP_c, \quad (10)$$

wherein $P_c$ is the static hardware power dissipated to operate the communication of a single user of the network. Based on (9) and (10), the system global energy efficiency (GEE) is defined as

$$\text{GEE}(K,\rho) = B\frac{K\mathbb{E}\left[R_1(K,\rho)\right]}{\rho + KP_c}, \quad (11)$$

which measures the amount of bits that can be reliably transmitted per Joule of consumed energy. Then, the GEE maximization problem is formulated as

$$\max_{\rho,K} \text{GEE}(K,\rho) \quad (12a)$$
$$\text{s.t. } 0 \leq \rho \leq \rho_{max} \quad (12b)$$
$$K \in \{K_{min}, K_{min}+1, \ldots, K_{max}\} \quad (12c)$$

The challenge posed by Problem (12) lies mainly in the presence of the integer variable $K$, which makes (12a) a non-differentiable function, and in its fractional nature, which makes (12) a non-convex problem even with respect to the continuous variable $\rho$. Moreover, available fractional programming tools require the numerator and denominator of the fractional objective to be concave and convex, respectively, in order to perform the maximization with polynomial complexity [3]. Instead, even allowing the integer variable $K$ to take on continuous values, the joint concavity of $KR(K,\rho)$ with respect to $(K,\rho)$ is not clear.

The approach to be used in the rest of this work will be based on a suitable combination of fractional programming theory and integer relaxation methods and two algorithms will be developed. The former will globally solve (12), while the latter is not provably convergent but enjoys an extremely limited computational complexity.

#### B. Proposed approach

To begin with, we will study the properties of (12) with respect to one variable at a time, determining the optimal $\rho$, for any fixed $K$, and the optimal $K$, for any fixed $\rho$. Based on these results, the optimal pair $(K,\rho)$ will be derived.

*1) Optimal $\rho$ for fixed $K$:* If $K$ is fixed, (12) becomes a single-ratio fractional problem, whose objective function is not concave, thus preventing the direct use of standard convex optimization methods. Despite this, the optimal $\rho$ can be determined in closed-form as shown next.

**Proposition 2.** *For any given $K$, the corresponding optimal $\rho$ that globally solves Problem* (12) *is given by*

$$\rho^* = \min(\rho_{max}, \rho_s), \quad (13)$$

*with $\rho_s$ being the unique stationary point of* (12a) *with respect to $\rho$.*

*Proof:* The objective (12a) is non-negative. Moreover, the argument of the statistical expectation at the numerator of (12a) tends to zero for $\rho \to 0$, while it grows logarithmically for $\rho \to \infty$. Since this holds for any realization of the random variable to be averaged, we obtain that (12a) tends to zero both when $\rho \to 0$ and $\rho \to \infty$. Thus, for any given $K$, (12a) must have a finite maximizer with respect to $\rho$.

Next, we observe that with respect to $\rho$, the function $R(K,\rho)$ is strictly concave, because the logarithm is strictly concave and the statistical average is a linear operator, thus preserving the concavity property of its argument. Thus, (12a) has a strictly concave numerator and an affine denominator with respect to $\rho$, and this implies that it is a strictly pseudo-concave function. In turn, the strict pseudo-concavity directly implies that (12a) has only one stationary point, $\rho_s$, which is also the function global maximizer. Thus, (12a) is a unimodal function of $\rho$, being increasing for $\rho \leq \rho_s$, decreasing for $\rho \geq \rho_s$, and with a peak for $\rho = \rho_s$. Finally, the thesis follows considering also (12b). ∎

*2) Optimal $K$ for fixed $\rho$:* The analysis of (12) with respect to $K$ is more involved, due to the fact that $K$ is an integer variable. Typically, exhaustive searches over the feasible set are employed to optimize integer variables. However, this might be time-consuming, especially for large $K_{max}$ and low $K_{min}$. Luckily, we will show how the optimal integer $K$ can be determined for any fixed $\rho$ without performing any exhaustive search. The approach will be to first relax $K$ to a continuous variable, allowing it to take values in interval of the real line $[K_{min}, K_{max}]$. Then, based on the continuous solution, the optimal integer $K$ in the discrete set

$\{K_{min},\ldots,K_{max}\}$ will be determined. We start our analysis showing the following lemma.

**Lemma 1.** *Consider the following function of two real variables:*

$$g : (K,\rho) \in \mathbb{R}_0^+ \times \mathbb{R}_0^+ \to g(K,\rho) = KR(K,\rho) \ . \quad (14)$$

*The function g is jointly strictly concave in K and $\rho$.*

*Proof:* With respect to $\rho$, (14) is equal to the function $KR(K,\rho)$ at the numerator of (12a), which has been already proved to be strictly concave for any fixed $K$. Next, as a function of both $K$ and $\rho$, $g(K,\rho)$ can be seen to be the perspective function of $R(\cdot,\rho)$. This directly implies the thesis, since the perspective operation preserves concavity [31]. ∎

At this point, let us consider the relaxed version of Problem (12), namely

$$\max_{K,\rho} \frac{g(K,\rho)}{\rho + KP_c} \quad (15a)$$

$$\text{s.t. } K \in [K_{min}, K_{max}] \quad (15b)$$

$$\rho \in [0, \rho_{max}] \quad (15c)$$

Lemma 1 enables to characterize, for any fixed $\rho$, the optimal $K \in [K_{min}, K_{max}]$, as well as the optimal integer value of $K \in \{K_{min}, K_{max}\}$, as shown in Proposition 3.

**Proposition 3.** *Assume $\rho$ is fixed in both (12) and its relaxed version (15). Then, the global solution of the relaxed problem (15) corresponding to the fixed value of $\rho$ is given by*

$$\bar{K} = \max(K_{min}, \min(K_{max}, K_s)) \ , \quad (16)$$

*with $K_s$ the unique stationary point of the function g with respect to K. Moreover, the optimal integer K that globally solves Problem (12) is given by*

$$K^* = \arg\max_{K \in \{\lceil \bar{K} \rceil, \lfloor \bar{K} \rfloor\}} \text{GEE}(K,\rho) \ , \quad (17)$$

*Proof:* The proof is omitted due to space constraints. It can be found in [30]. ∎

*3) Joint optimization of $\rho$ and $K$:* After performing the separate optimization of $\rho$ and $K$, this section tackles the joint optimization of both variables $\rho$ and $K$. Specifically, two provably convergent algorithms will be devised. The former will achieve the global optimum of (12), while the latter will trade-off optimality with a lower computational complexity.

The first approach is based on the consideration that for any fixed $K$, the optimal transmit power has been characterized in (13), which only requires to compute the unique stationary point $\rho_s$ of (12a). Thus, for any given $K$, the corresponding optimal $\rho$ can be computed with a negligible complexity, by just solving the scalar equation:

$$\frac{\partial \text{GEE}(K,\rho)}{\partial \rho} = 0 \ , \quad (18)$$

which is guaranteed to admit a unique solution. Therefore, (12a) can be globally solved by computing, for any feasible $K \in \{K_{min},\ldots,K_{max}\}$, the corresponding optimal $\rho$ and then picking the best pair $(K,\rho)$. The formal procedure is stated as in Algorithm 1.

It should be stressed that although Algorithm 1 checks all $K \in \{K_{min},\ldots,K_{max}\}$, the resulting complexity is still affordable. Indeed, for each value of $K$, no numerical

---

**Algorithm 1** Global optimization of $(K,\rho)$

**for** $K = K_{min}{:}K_{max}$ **do**
  Set $K_s$ as the unique solution of (18);
  Set $\bar{\rho}_K$ as in (13);
**end for**
$(K,\rho) = \arg\max_{(\bar{\rho}_K, K)} \text{GEE}(K,\rho);$

---

optimization needs to be performed, but rather only a scalar equation must be solved. This can be accomplished with limited complexity by solving all equations in parallel and off-line employing standard methods (e.g. bisection or Newton's method). Moreover, the overall complexity can be reduced by the following corollary of Proposition 2, which provides a way to restrict the range in which the solution of (18) will lie for a given $K$, based on the solution obtained for the previous value of $K$.

**Corollary 1.** *Let $K_\ell \in \{K_{min},\ldots,K_{max}\}$, and denote by $\rho_{s,\ell}$ and $\rho_{s,\ell+1}$ the solutions of (18) corresponding to $K = K_\ell$ and $K_{\ell+1}$. Then it holds*

$$\rho_{s,\ell+1} \in [\rho_\ell, \rho_{max}], \text{ if } \frac{\partial \text{GEE}(\rho_{s,\ell}, K_{\ell+1})}{\partial \rho} \geq 0 \quad (19)$$

$$\rho_{s,\ell+1} \in [0, \rho_\ell], \text{ if } \frac{\partial \text{GEE}(\rho_{s,\ell}, K_{\ell+1})}{\partial \rho} \leq 0 \quad (20)$$

*Proof:* From Proposition 2, we know that $\text{GEE}(K,\rho)$ is a unimodal function of $\rho$ with a unique stationary point for any given $K$. Thus, $\rho_{s,\ell+1}$ is the unique solution of (18) when $K = K_{\ell+1}$, and it holds that

$$\frac{\partial \text{GEE}(\rho, K_{\ell+1})}{\partial K} \geq 0 \ , \ \forall \ K \leq K_{s,\ell+1} \quad (21)$$

$$\frac{\partial \text{GEE}(\rho, K_{\ell+1})}{\partial \rho} \leq 0 \ , \ \forall \ \rho \geq \rho_{s,\ell+1} \quad (22)$$

Hence, if the derivative of the GEE function evaluated at $(\rho_{s,\ell}, K_{\ell+1})$ is non-negative, then we are in the increasing part of the GEE function, and the unique stationary point $\rho_{s,\ell+1}$ will be larger than $\rho_{s,\ell}$. Hence (19) follows, while (20) can be obtained by a similar reasoning. Finally, according to (13) it holds

$$\rho^*_{\ell+1} = \min(\rho_{max}, \rho_{s,\ell+1}) \ , \quad (23)$$

which is an increasing function of $\rho_{s,\ell+1}$ ∎

The second algorithm to be developed, aims at having an even lower computational complexity, even if this comes at the price of possibly suboptimal performance. The approach is again based on the fact that for any $K$ we can easily find the corresponding optimal $\rho$ by (13). Moreover, for any $\rho$, we can find the optimal $K$ by (17). Again, this can be performed with a negligible complexity, since it requires only to find the unique solution of the scalar equation:

$$\frac{\partial \text{GEE}(K,\rho)}{\partial K} = 0 \ , \quad (24)$$

to determine the value of $K_s$. Therefore, a convenient approach to tackle (12a) is to employ the alternating optimization method, i.e. alternatively optimizing $\rho$ for fixed $K$, and $K$ for fixed $\rho$. The formal procedure is stated as in Algorithm 2 with $\text{GEE}^{(j)}$ being the achieved value of (12a) after iteration $j$.

Algorithm 2 is provably convergent in the value of the objective function. Indeed, for any $j$ it clearly holds $\text{GEE}^{(j)} \geq \text{GEE}^{(j-1)}$. Moreover, (12a) is a continuos function over a compact feasible set. Thus, by Weierstrass extreme value theorem, (12a) must be upper-bounded on the feasible set of Problem (12) and thus the sequence $\{\text{GEE}^{(j)}\}_j$ converges.

**Algorithm 2** Alternating optimization of $(K, \rho)$
---
Set $\varepsilon > 0$; Initialize $\rho$ to a feasible value;
**repeat**
   Given $\rho$ solve (24) and set $K$ as in (17);
   Given $K$ solve (18) and set $\rho$ as in (13);
**until** $|\text{GEE}^{(j)} - \text{GEE}^{(j-1)}| < \varepsilon$
---

## IV. NUMERICAL RESULTS

For the numerical performance analysis of the proposed algorithms, $P_c = 1$ Watt has been considered. In Fig. 2a, the maximum allowed power for all users is set to $\rho_{max} = 0\,\text{dBW}$. In this figure, the GEE versus $K$ is achieved by GEE maximization by Proposition 2 which is labeled as GEE and power allocation that maximizes the system sum-rate, i.e. $\rho = \rho_{max}$ which is labeled as $\text{GEE}_{max}$. As expected, it is seen that increasing the number of total transmit antennas $N_T$ results in better performance. The other point is the optimal GEE value coincides with the value obtained when the sum-rate is maximized for larger $K$. This is due to the fact that the GEE is a unimodal function of the transmit power with a finite maximizer. Moreover, increasing $K$ causes a larger static power consumption term $P_c$, which in turn leads to a larger optimal power level. Thus, as the antenna number gets larger, the optimal power level becomes not feasible for the considered value of $\rho_{max}$, and maximum power transmission is the optimal transmit policy. This circumstance is studied further in Fig. 2b, which considers a similar configuration, but with $\rho_{max} = 10\,\text{dBW}$. Unlike Fig. 2a, here it is seen that the GEE is constant with respect to the number of RF chains. Indeed, having a larger $\rho_{max}$ allows attaining the optimal power level which maximizes the GEE for any considered value of $K$.

In Fig. 3, we report the achieved GEE by jointly allocating the transmit power $\rho$ and the number of active RF chains $K$, according to the following schemes: 1) GEE maximization by Algorithm 1; 2) GEE maximization by Algorithm 2; 3) Resource allocation that maximizes the system sum-rate, i.e. $\rho = \rho_{max}$ and $K = K_{max}$. Remarkably, the results indicate that Algorithm 1 and Algorithm 2 achieve the same performance, despite the fact that Algorithm 2 has a lower complexity than Algorithm 1. Moreover, the GEE achieved by rate optimization performs worse than other schemes, thus confirming that the GEE is not monotonically increasing in $\rho$ and $K$, but instead has a finite maximizer ($\rho^* \neq \rho_{max}, K^* \neq K_{max}$). Moreover, the figure confirms that the GEE as a function of $\rho$ is first increasing and then decreasing.

## V. CONCLUSION

Two convergent and computationally-affordable algorithms for GEE maximization in mmWaves, large-scale MIMO systems with spatial modulation and hybrid-beamforming have been proposed. The two algorithms strike different complexity-optimality trade-offs, with the former being provably optimal, and the latter enjoying a very limited complexity, although being possibly suboptimal. Nevertheless, numerical results show that the two algorithms perform very similarly.

REFERENCES

[1] Ericsson White Paper, "More than 50 billion connected devices," Ericsson, Tech. Rep. 284 23-3149 Uen, Feb. 2011.
[2] "NGMN alliance 5G white paper," *https://www.ngmn.org/5g-white-paper/5g-white-paper.html*, 2015.
[3] A. Zappone and E. Jorswieck, "Energy efficiency in wireless networks via fractional programming theory," *Foundations and Trends® in Communications and Information Theory*, vol. 11, no. 3-4, pp. 182–396, 2007. [Online]. Available: http://dx.doi.org/10.1561/0100000088.
[4] S. Buzzi *et al.*, "A survey of energy-efficient techniques for 5G networks and challenges ahead," *IEEE Journal on Selected Areas in Communications*, vol. 34, no. 4, pp. 697–709, Apr. 2016.
[5] D. W. K. Ng, E. S. Lo, and R. Schober, "Energy-efficient resource allocation in OFDMA systems with large numbers of base station antennas," *IEEE Transactions on Wireless Communications*, vol. 11, no. 9, pp. 3292–3304, Sep. 2012. DOI: 10.1109/TWC.2012.072512.111850.
[6] G. Miao, N. Himayat, G. Y. Li, and S. Talwar, "Distributed interference-aware energy-efficient power optimization," *IEEE Transactions on Wireless Communications*, vol. 10, no. 4, pp. 1323–1333, Apr. 2011.
[7] F. Rusek, D. Persson, B. K. Lau, E. G. Larsson, T. L. Marzetta, O. Edfors, and F. Tufvesson, "Scaling up MIMO: Opportunities and challenges with very large arrays," *IEEE Signal Processing Magazine*, vol. 30, no. 1, pp. 40–60, 2013.
[8] T. L. Marzetta, "Massive MIMO: An introduction," *Bell Labs Journal*, vol. 20, pp. 11–22, 2015.
[9] J. Hoydis, S. ten Brink, and M. Debbah, "Massive MIMO in the UL/DL of cellular networks: How many antennas do we need?" *IEEE Journal on Selected Areas in Communications*, vol. 31, no. 2, pp. 160–171, Feb. 2013.
[10] T. S. Rappaport, S. Sun, R. Mayzus, H. Zhao, Y. Azar, K. Wang, G. N. Wong, J. K. Schulz, Samimi, and F. Gutierrez, "Millimeter wave mobile communications for 5G cellular: It will work!" *IEEE Acess*, vol. 1, pp. 335–349, 2013.
[11] T. S. Rappaport, R. W. Heath Jr., J. N. Murdock, and R. C. Daniels, *Millimeter Wave Wireless Communications*. Pearson, 2014.
[12] T. Kim, J. Park, J.-Y. Seol, S. Jeong, J. Cho, and W. Roh, "Tens of Gbps support with mmwave beamforming systems for next generation communications," in *IEEE Global Communications Conference*, Dec. 2013, pp. 3685–3690.
[13] O. E. Ayach, S. Rajagopal, S. Abu-Surra, Z. Pi, and R. W. Heath Jr., "Spatially sparse precoding in millimeter wave MIMO systems," *IEEE Transactions on Wireless Communications*, vol. 13, no. 3, pp. 1499–1513, Mar. 2014.
[14] R. Méndez-Rial, C. Rusu, N. González-Prelcic, A. Alkhateeb, and R. W. H. Jr., "Hybrid MIMO architectures for millimeter wave communications: Phase shifters of switches?" *IEEE Access*, vol. 4, pp. 247–267, 2016.
[15] A. Alkhateeb, G. Leus, and R. W. Heath Jr., "Limited feedback hybrid precoding for multi-user millimeter wave systems," *IEEE Transactions on Wireless Communications*, vol. 14, no. 11, pp. 6481–6494, Nov. 2015.
[16] F. Sohrabi and W. Yu, "Hybrid digital and analog beamforming design for large-scale MIMO systems," in *IEEE International Conference on Acoustics, Speech and Signal Processing*, Apr. 2015, pp. 2929–2933.
[17] S. Payami, M. Ghoraishi, and M. Dianati, "Hybrid beamforming for large antenna arrays with phase shifter selection," *IEEE Transactions on Wireless Communications*, vol. 15, no. 11, pp. 7258–7271, Nov. 2016.
[18] R. Y. Mesleh, H. Haas, S. Sinanovic, C. Wook, and S. Yun, "Spatial modulation," *IEEE Transactions on Vehicular Technology*, vol. 57, no. 4, pp. 2228–2241, Jul. 2008.
[19] M. Di Renzo, H. Haas, and P. Grant, "Spatial modulation for multiple-antenna wireless systems: A survey," *IEEE Communications Magazine*, vol. 49, no. 12, pp. 182–191, 2011.

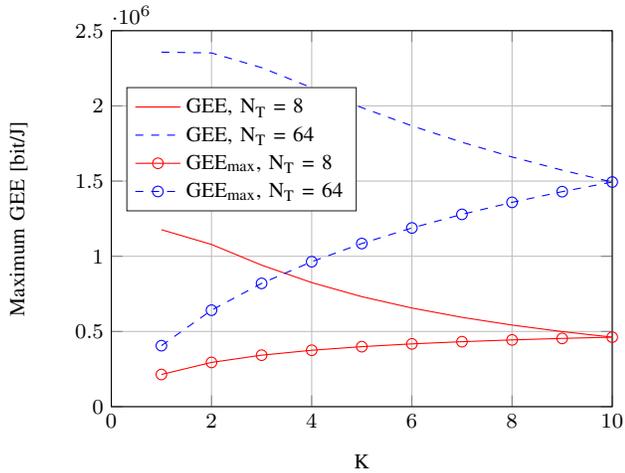
(a) GEE maximization with $\rho_{max} = 0$dBW

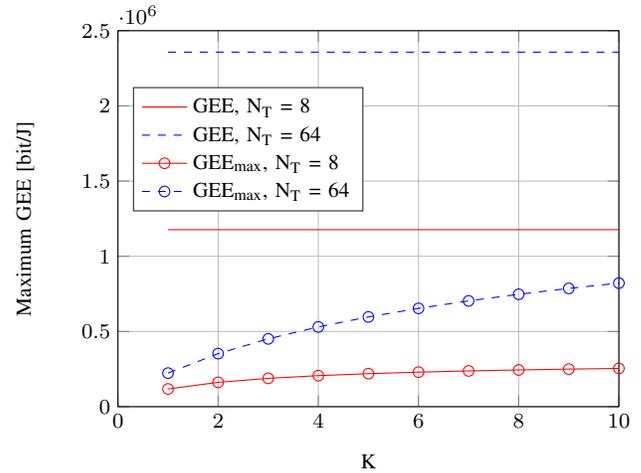
(b) GEE maximization with $\rho_{max} = 10$dBW

Fig. 2: Optimization of $\rho$ for fixed $K$ with $N_A = 4, N_R = 1, P_c = 1$ Watt, $4-QAM$

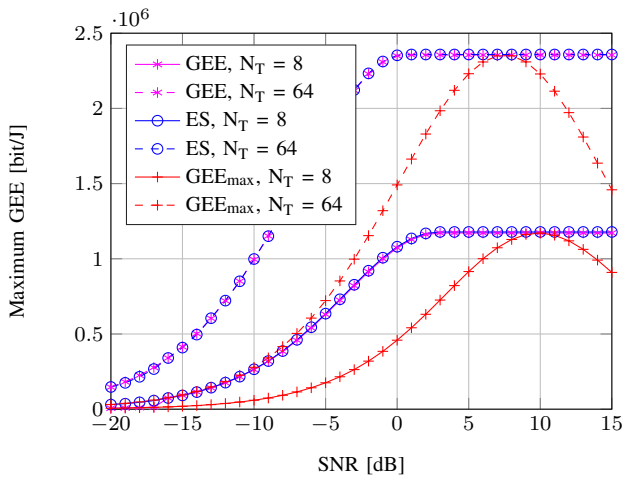

Fig. 3: Joint optimization of $(K, \rho)$ with $N_A = 4, N_R = 1, P_c = 1$ Watt, $4 - QAM$

[20] M. D. Renzo, H. Haas, A. Ghrayeb, S. Sugiura, and L. Hanzo, "Spatial modulation for generalized MIMO: Challenges, opportunities and implementation," *IEEE in Proceedings*, vol. 102, no. 1, pp. 56–103, Jan. 2014.

[21] M.-C. Lee, W.-H. Chung, and T.-S. Lee, "Generalized precoder design formulation and iterative algorithm for spatial modulation in MIMO systems with CSIT," *IEEE Transactions on Communications*, vol. 63, no. 4, pp. 1230–1244, Apr. 2015.

[22] S. Narayanan, M. J. Chaudhry, A. Stavridis, M. D. Renzo, F. Graziosi, and H. Haas, "Multi-user spatial modulation MIMO," in *IEEE Wireless Communications and Networking Conference*, Apr. 2014, pp. 671–676.

[23] N. Ishikawa, R. Rajashekar, S. Sugiura, and L. Hanzo, "Generalized-spatial-modulation-based reduced-RF-chain millimeter-wave communications," *IEEE Transactions on Vehicular Technology*, vol. 66, no. 1, pp. 879–883, Jan. 2017.

[24] M.-C. Lee and W.-H. Chun, "Transmitter design for analog beamforming aided spatial modulation in millimeter wave MIMO systems," in *IEEE International Symposium on Personal, Indoor and Mobile Radio Communications: Fundamentals and PHY*, Sep. 2016, pp. 1–6.

[25] Y. Cui, X. Fang, and L. Yan, "Hybrid spatial modulation beamforming for mmwave railway communication systems," *IEEE Transactions on Vehicular Technology*, vol. 65, no. 12, pp. 9597–9606, Dec. 2016.

[26] E. Björnson, L. Sanguinetti, J. Hoydis, and M. Debbah, "Optimal design of energy-efficient multi-user MIMO systems: Is massive MIMO the answer?" *IEEE Transactions on Wireless Communications*, vol. 14, no. 6, Jun. 2015.

[27] A. Pizzo and L. Sanguinetti, "Optimal design of energy-efficient millimeter wave hybrid transceivers for wireless backhaul," in *15th International Symposium on Modelling and Optimization in Mobile, Ad Hoc, and Wireless Networks (WiOpt)*, 2017, pp. 1–8.

[28] A. A. M. Saleh and R. A. Valenzuela, "A statistical model for indoor multipath propogation," *IEEE Journal on Selected Areas in Communications*, vol. SAC-5, no. 2, pp. 128–137, Feb. 1987.

[29] M. Yüzgeçcioğlu and E. Jorswieck, "Hybrid beamforming with spatial modulation in multi-user massive MIMO mmWave networks," in *IEEE International Symposium on Personal, Indoor, and Mobile Radio Communications (PIMRC)*, Oct. 2017.

[30] H.-G. Engler, M. Yüzgeçcioğlu, A. Zappone, and E. A. Jorswieck, "Energy efficiency optimization in mmWave massive mimo systems with hybrid beamforming," to be submitted 2018.

[31] S. Boyd and L. Vandenberghe, *Convex Optimization*. Cambridge University Press, 2004.